# Bubble growth on arrays of micro-electrodes

Mengyuan Huang[1,2], Gerd Mutschke[2✉], Zhicheng Yuan[1✉] and Kerstin Eckert[2,3]

*1. College of Automotive and Energy Engineering, Tongji University, Shanghai, 201804, China*

*2. Institute of Fluid Dynamics, Helmholtz-Zentrum Dresden-Rossendorf, Dresden, 01328, Germany*

*3. Institute of Process Engineering and Environmental Technology, Technical University of Dresden, Dresden, 01069, Germany*

✉ zhicheng_yuan@tongji.edu.cn, g.mutschke@hzdr.de

**Abstract**

Gas bubbles evolving on electrodes during water-electrolysis are blocking active reaction area, thus hindering mass transfer and raising Ohmic resistance. Unlike earlier models that prescribe a uniform current density on the wetted part of the electrode, we resolve the primary electric field, which allows the current density and the interfacial gas production to respond to the geometry of the electrode and the temporal evolution of the bubbles. Using three-dimensional geometrical volume-of-fluid (VOF) simulations with phase change in Basilisk, we examine the growth of single-bubbles on electrodes of different size and of multiple bubbles growing on arrays of catalytic electrode islands. The non-uniform current density and the associated Ohmic resistance significantly affect the growth dynamics. Unlike the case of a single bubble, the outer bubbles in case of electrode islands tend to drift outward during growth, thus delaying full electrode coverage and sustaining current. Footprint tracking and a theoretical analysis show that this drift is governed by the liquid advection driven by the growth of neighboring bubbles, scaling with their separation $\sim 1/d^2$, and modulated by the current-density asymmetry. These results show how electrode patterning and bubble spacing can be exploited to tailor the electric field distribution and reduce bubble-induced resistive losses during water electrolysis.

**Nomenclature**

| | |
|---|---|
| $A_e$ | electrode area ($m^2$) |
| $c$ | dissolved gas concentration ($mol/m^3$) |
| $D$ | diffusion coefficient of dissolved gas ($m^2/s$) |
| $d$ | distance between neighboring electrode-bubble pairs (m) |
| $d_e$ | electrode diameter (m) |
| $F$ | Faraday constant (C/mol) |
| $g$ | gravitational acceleration ($m/s^2$) |
| $I_{cell}$ | cell current (A) |
| $j$ | current density ($A/m^2$) |
| $L_0$ | edge length of the computational domain (m) |
| $\dot{m}$ | interfacial mass flux ($kg/(m^2 \cdot s)$) |
| $M_g$ | molar mass of the gas (kg/mol) |
| $n$ | number of bubbles |
| $p$ | pressure (Pa) |
| $R$ | Ohmic resistance (Ω) |
| $r_b$ | bubble radius (m) |
| $r_e$ | electrode radius (m) |
| $t$ | time (s) |
| $\boldsymbol{U}$ | velocity vector (m/s) |
| $E$ | cell voltage (V) |
| $u_d$ | drift velocity (m/s) |
| $V_b$ | bubble volume (m³) |
| $z$ | charge transfer number |
| $\alpha_l$ | volume fraction |
| $\gamma$ | surface tension coefficient (N/m) |
| $\eta_{Faraday}$ | Faraday efficiency |
| $\kappa$ | electric conductivity (S/m) |
| $\mu$ | dynamic viscosity (Pa·s) |
| $\rho$ | density (kg/m³) |
| $\varphi_e$ | electric potential (V) |
| *subscripts l, g* | liquid phase, gas phase |

## 1 Introduction

Green hydrogen produced by water electrolysis is a key technology in the development of a clean energy system, and improving the efficiency of electrolyzers is of major economic importance. A considerable part of the energy losses during electrolysis originates from the gas bubbles that nucleate and grow at the electrodes: attached bubbles block electro-active area, hinder the transfer of dissolved gas and force the electric current to bypass the non-conducting gas phase, thereby modifying the electric field and raising the resistance of the cell (Kempler et al. 2024). The electric field, in turn, determines the local gas production on the electrode and the associated bubble growth. A detailed understanding of the interaction between the bubble dynamics and the electric field is essential for designing electrodes to reduce bubble-related energy losses (Bashkatov et al. 2022; Rox et al. 2025).

In recent years, numerical simulations have become a powerful tool to study electrolytic bubbles, as they give access to quantities that are difficult to measure, such as the local current density, the concentration of dissolved gas or ionic species near the bubbles (Vachaparambil and Einarsrud 2021). Interface-resolving simulations have addressed, among others, the oversaturation-driven growth of surface bubbles (Gennari et al. 2022), the influence of buoyancy-driven convection on the gas transport at electrodes (Sepahi et al. 2024), the Marangoni flow around growing bubbles (Massing et al. 2019; Meulenbroek et al. 2023; Khalighi et al. 2025) and the coalescence between bubbles (Bashkatov et al. 2024; Qin et al. 2026). Among different codes, Basilisk has recently drawn attention for simulating electrolytic bubbles. It possesses high accuracy in phase-change problems thanks to the geometrical volume-of fluid (VOF) method combining height function and a balanced-force discretization scheme to calculate the surface tension with an effective octree-based adaptive mesh refinement. However, in most of these studies, the electric field is not resolved. The electrochemical reaction is modelled only through a prescribed, uniform current density, or equivalently a fixed interfacial gas flux based on Faraday's law (Gennari et al. 2022; Huang et al. 2025; Qin et al. 2026).

In reality, the current density on the electrode is inherently non-uniform: the current lines have to bypass the attached bubbles, they concentrate near the electrode edges, and this distribution changes continuously during the bubble evolution. When growing bubbles progressively cover the electrode, the non-uniform distribution of the current density and the local gas production must respond to this time-dependent geometry, which cannot be captured by the constant-current-density assumption. Resolving the electric field alongside the two-phase flow is thus essential for a complete, two-way coupling between the bubble dynamics and the electrochemistry.

The distribution of the electric field is also closely tied to the structure of the electrode. Electrodes patterned with small catalytic islands or spots are known to be beneficial for the overall electrolysis efficiency, as they allow control over the nucleation sites, the electrocatalytic area, the wetting behavior and the bubble size (Heinrich et al. 2024; Rox et al. 2025). On such electrodes, multiple bubbles grow simultaneously in close proximity, and they could interact through the shared electric field, the dissolved gas field and the liquid flow induced during their growth. To derive the optimal spacing of the catalytic islands, it is essential to understand how these interactions may affect the coverage of the electrode and the cell current.

In this work, we present three-dimensional simulations of bubble growth during water electrolysis in which the primary electric field is resolved and fully coupled to the interfacial gas production and the bubble dynamics. The simulations are performed with the open-source software Basilisk (Popinet 2014) using a geometrical VOF method with adaptive mesh refinement (Popinet 2009), combined with the phase-change model of Gennari et al. (2022) for the interfacial mass transfer. Starting from a single bubble on electrodes of different size, we proceed to arrays of bubbles growing on an electrode composed of small catalytic islands, thereby combining numerical and theoretical studies. The remainder of the paper is organized as follows: Section 2 introduces the numerical model and the simulation setup, Section 3 first presents a validation of the numerical model using analytical solutions of the primary resistance in absence of bubbles, then the numerical results for a single bubble and for bubble arrays supported by a theoretical analysis of growth-driven advection between neighboring bubbles. Section 4 finally summarizes the conclusions.

## 2 Numerical model and simulation setup

### 2.1 Governing equations

The two-phase flow of electrolyte and gas is described by the incompressible Navier–Stokes equations, Eq. (1):

$$\frac{\partial(\rho \boldsymbol{U})}{\partial t} + \nabla \cdot (\rho \boldsymbol{U}\boldsymbol{U}) = -\nabla p + \nabla \cdot \{\mu(\nabla \boldsymbol{U} + (\nabla \boldsymbol{U})^T)\} + \boldsymbol{f}_\gamma + \rho \boldsymbol{g}, \quad \nabla \cdot \boldsymbol{U} = \dot{S}_\rho, \qquad \text{Eq. (1)}$$

with $U, p, \rho, \mu$ and $\boldsymbol{g}$ denoting the phase-averaged velocity, the pressure, the density, the dynamic viscosity and the gravitational acceleration, respectively. The gas-liquid interface is captured using a geometrical VOF method:

$$\frac{\partial \alpha_l}{\partial t} + \nabla \cdot (\alpha_l \mathbf{U}) = \dot{S}_\alpha, \qquad \dot{S}_\alpha = \frac{\dot{m}}{\rho_l}\,\delta_\Sigma \qquad \text{Eq. (2)}$$

Here, $\alpha_l$ denotes the liquid volume fraction that equals to 1 in the liquid phase and 0 in the gas phase. $\dot{S}_\alpha$ is the source term caused by the phase change, with $\delta_\Sigma$ being a Dirac function that has a non-zero value only at the interface, and $\dot{m}$ being the interfacial mass transfer rate that will be introduced later in Eq. (7). The surface tension force $\boldsymbol{f}_\gamma$ is included as a continuum body force acting at the interface, Eq. (3), and the material properties are averaged between the liquid ($l$) and the gas ($g$) phases according to the local volume fraction, Eq. (4):

$$\boldsymbol{f}_\gamma = \gamma\,\kappa_\Sigma\,\boldsymbol{n}_\Sigma\,\delta_\Sigma. \qquad \text{Eq. (3)}$$

$$\rho = \alpha_l \rho_l + (1-\alpha_l)\rho_g, \qquad \mu = \alpha_l \mu_l + (1-\alpha_l)\mu_g. \qquad \text{Eq. (4)}$$

Here, $\gamma$ represents the surface tension coefficient, $\kappa_\Sigma$ and $\boldsymbol{n}_\Sigma$ the curvature and unit normal of the interface, respectively. Further, the phase change at the interface gives rise to a non-zero velocity divergence, Eq. (5):

$$\dot{S}_\rho = \dot{m}\left(\frac{1}{\rho_g} - \frac{1}{\rho_l}\right)\delta_\Sigma, \qquad \text{Eq. (5)}$$

The transport of the dissolved gas in the electrolyte is governed by an advection-diffusion equation with a sink term at the bubble interface, Eq. (6):

$$\frac{\partial c}{\partial t} + \nabla \cdot (\mathbf{U}c) = \nabla \cdot (D\nabla c) + \dot{S}_c, \;\; \dot{S}_c = -\frac{\dot{m}}{M_g}\,\delta_\Sigma \qquad \text{Eq. (6)}$$

Following the phase-change model of Gennari et al. (2022), the interfacial mass flux is computed from the concentration gradient in the liquid side of the interface based on the Fick's law, Eq. (7), such that the bubble growth is driven by the local supersaturation of dissolved gas. This model has been validated against Scriven and Epstein-Plesset solutions (Gennari et al. 2022), as well as experimental results (Huang et al. 2026) for the oversaturation-driven growth of surface bubbles.

$$\dot{m} = \frac{-M_g D}{1 - \rho_g/\rho} \frac{\partial c}{\partial \boldsymbol{n}_\Sigma}, \qquad \text{Eq. (7)}$$

with $M_g, D$ being the molar mass and diffusion coefficient of the gas, and $\boldsymbol{n}_\Sigma$ is the unit normal at the interface. The gas production at the wetted electrode surface is modelled as a Neumann boundary condition of $c$ based on Faraday's law, as detailed in Section 2.2.

Assuming electro-neutrality and constant electric conductivity in both phases, and neglecting the Butler-Volmer electrode kinetics, the electric field $\varphi_e$ can be solved by a Poisson equation based on the charge conservation:

$$\boldsymbol{j} = -\kappa \nabla \varphi_e, \qquad \nabla \cdot \boldsymbol{j} = 0, \qquad \text{Eq. (8)}$$

with $\kappa$ being a phase-dependent conductivity:

$$\kappa = \frac{1}{\frac{\alpha_l}{\kappa_l} + \frac{\alpha_g}{\kappa_g}}. \qquad \text{Eq. (9)}$$

Here, a harmonic average weighted by the local volume fractions is applied, as it shows better numerical stability in our simulations. Since the conductivity of the electrolyte is set significantly larger than that of the gas, the current lines are deflected around the bubbles. This causes a non-uniform current density, in contrast to the uniform-current-density assumption commonly made in earlier studies, where the Poisson equation for the electric potential (8) is not solved, and only a constant flux is assumed at the bubble foot. However, these details do affect the bubble growth behavior. The local current density obtained from the electric field determines the interfacial gas production via Faraday's law. The governing equations are discretized and solved with the finite-volume framework Basilisk on adaptive octree grids (Popinet 2009, 2014).

## 2.2 Computational domain and boundary conditions

The computational domain is a cube of edge length $L_0 = 0.01$ m, as sketched in Fig. 1. The working electrode (cathode) is located at the bottom wall, and the counter-electrode (anode) is the entire top wall, with a fixed potential difference applied between them. The octree grid is adaptively refined based on the interface position and on the gradients of the velocity, the concentration and the electric potential, with refinement levels between 4 and 9, corresponding to a minimum cell size of about 20 μm.

In the single-bubble cases, two electrode configurations are compared: either the entire bottom wall is electro-active (case L-e), or a small disk electrode of 1.5 mm in radius centered at the bottom of the domain is considered, surrounded by insulating Teflon (case S-e). Initially, a spherical bubble with a contact radius of 1.2 mm and a water-side contact angle of 45° sits at the bottom center. This results in an electrode area of either 100 mm$^2$ (L-e) or 7.07 mm$^2$ (S-e), and an initial bubble contact area of 4.52 mm$^2$. The S-e case is then extended to multiple-bubble cases (M-e) by considering an electrode patterned with several circular electro-active disks embedded in insulating Teflon. Hereby, hemispherical bubbles are placed at the center of each disk. The geometry is chosen such that initially the total electrode area equals to 9.42 mm$^2$, and the total de-wetted electrode area equals to 6.03 mm$^2$ in cases with different electrode-bubble pairs (see Fig. 6). The center-to-center distance between neighboring electrodes is 2.4 $r_e$ by default, where $r_e$ denotes the radius of one electro-active spot, and it is later varied to investigate its influence on the bubble behavior. The smallest and largest distance considered in this work is 2.1 $r_e$ and 4.2 $r_e$, respectively.

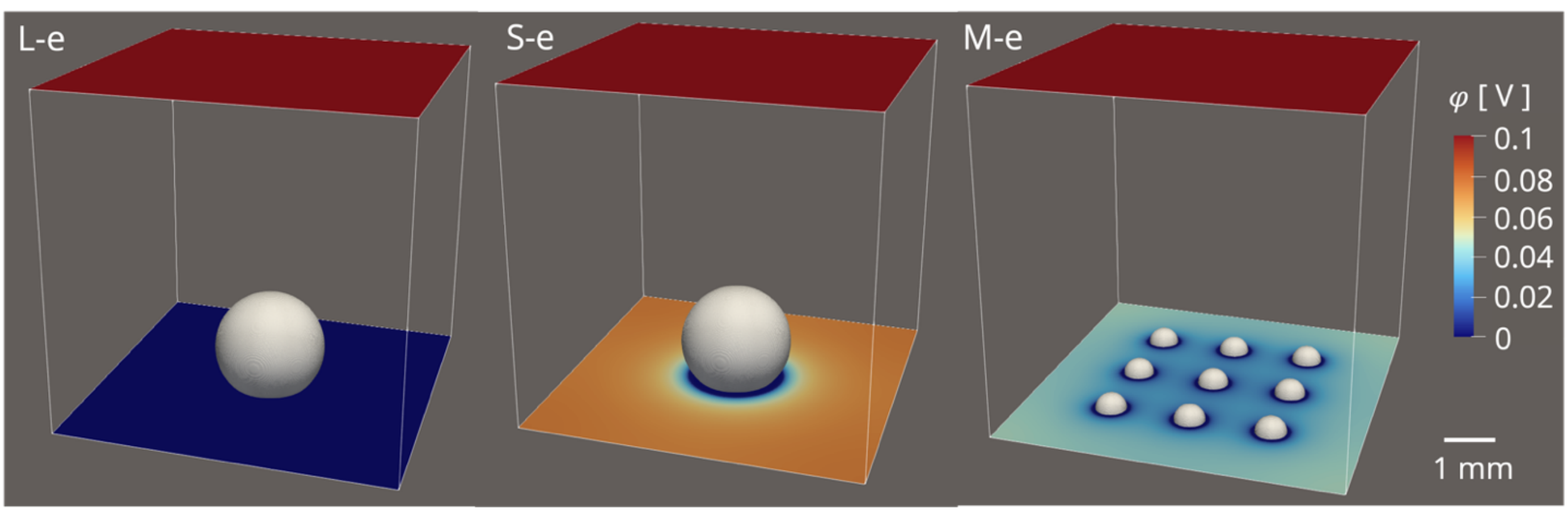

**Fig. 1**. The computational domain for the single-bubble cases where the entire bottom (L-e) or a small disk around the bubble (S-e) is the cathode, and for the multiple-bubble case (M-e) with nine bubble-electrode pairs. The color surface shows the initial electric potential distribution on the anode (top) and the cathode (bottom).

If not otherwise stated, the cell voltage is kept fixed at 0.1 V, resulting in a cell current of a few milliampere depending on the geometry of the electrode. At the bottom wall, a no-slip condition is applied for the velocity, and a static contact angle of 90° is prescribed. At the electro-active area, the flux of dissolved gas is obtained from the computed local current density via Faraday's law, Eq. (10), with $z$, $F$ and $D$ being the charge transfer number, the Faraday constant and the diffusion coefficient of the dissolved gas. The insulating parts of the bottom wall are treated as no-flux boundaries for both the dissolved gas and the electric potential.

$$\frac{\partial c}{\partial n} = \frac{j}{zFD} \quad \text{Eq. (10)}$$

For numerical reasons, the electric conductivity is set to be 0.1 S/m in the gas phase and 10 S/m in the liquid phase. As shown in Fig. A1 in the Appendix, the chosen $\kappa_g$ is sufficiently small to ensure accurate results. Other material properties are same as in Huang et al. (2025).

## 3 Results

### 3.1 Single bubble on a single electrode

#### 3.1.1 Validation of the electric field implementation

For bubble-free electrodes, classical solutions of the primary electric field are available, e.g. for a small disk electrode embedded in an insulating plane (Newman 1966). In the following, we validate our implementation of the primary electric field using available theoretical solutions of the primary resistance $R = E/I$, with $E$ and $I$ being the cell voltage and the cell current.

In absence of bubbles, the case L-e can be regarded as a bulk resistor:

$$R_{bulk} = \frac{L_0}{A_e \kappa_l}, \quad \text{Eq. (11)}$$

with $L_0, A_e$ and $\kappa_l$ being the distance between the counter- and working electrodes, the electrode area, and the electric conductivity in the liquid.

For the case where the counter-electrode is infinitely large and infinitely far away from a small disk working electrode, an analytical solution was derived by Newman (1966):

$$R_{Newman} = \frac{1}{2\kappa_l d_e}, \quad \text{Eq. (12)}$$

with $d_e$ being the electrode diameter. Our case S-e without bubble is expected to converge to this solution if the cathode is much smaller than the anode ($d_e \ll L_0$).

As can be seen in Fig. 2, the simulated primary electric resistance in absence of bubble agrees well with the theoretical solutions. In the L-e case ($d_e \to L_0$), $R_{sim}$ equals to $R_{bulk}$, and is two times larger than $R_{Newman}$, as in a cubic domain $A_e = L_0^2 \approx d_e^2$. When $d_e = 0.5L_0$, $R_{Newman} = R_{bulk}$ based on their definitions, both being slightly smaller than $R_{sim}$. When the electrode size further shrinks, the simulated resistance quickly increases, becoming significantly larger than $R_{bulk}$, and converges towards $R_{Newman}$. The agreement between numerical and theoretical solutions shows the validity of the electric field calculations. We remark that the S-e case corresponds to $d_e = 0.3L_0$, where $R_{bulk} = 10\ \Omega$, $R_{Newman} = 16.7\ \Omega$ and $R_{sim} = 20.7\ \Omega$ without bubble. If there is an initial bubble, $R_{sim}$ raises to 31.8 Ω due to the deflection of current lines around the bubble. These results show that the analytical solution could predict the order of magnitude of the Ohmic resistance for single-electrode cases, but it remains difficult to accurately calculate the resistance for cases where bubbles are involved.

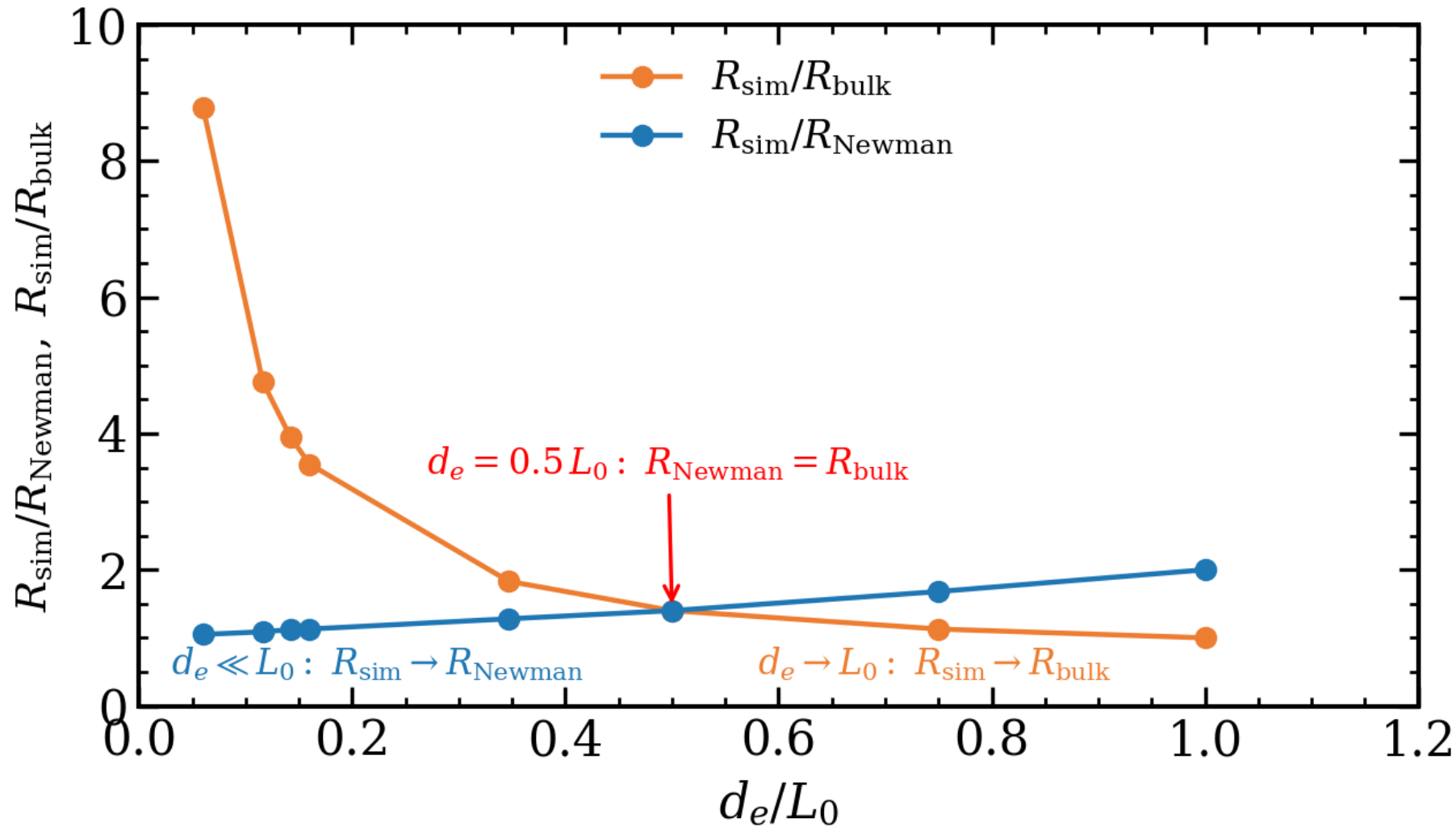

**Fig. 2**. Values of $R_{sim}/R_{bulk}$ and $R_{sim}/R_{Newman}$ depending on the ratio between the electrode diameter and the domain size. When $d_e \ll L_0$, the simulated resistance converges to $R_{Newman}$. When $d_e \to L_0$, the simulated resistance converges to $R_{bulk}$.

### 3.1.2 Non-uniform electric field and bubble growth

Next, we investigate how the calculated primary current density shapes the growth of a single bubble. In contrast to the commonly used approximation of a uniform current density at the electrode, the constant potential at the electrode yields a more realistic current density that varies strongly along the electrode, resulting in a non-uniform gas production.

Fig. 3 shows the distribution of the current density and of the dissolved gas concentration in the L-e case at 20 s. Due to the shielding of the bubble, the current density reduces near the bubble foot, but approaches a constant value on the electrode far away from the bubble. Around the equator of the bubble, the current density is slightly enhanced as the current lines are squeezed due to the narrowing of the liquid region. The dissolved gas concentration reduces near the bubble foot, according to both, a smaller current density and the ongoing mass transfer of dissolved gas into the bubble.

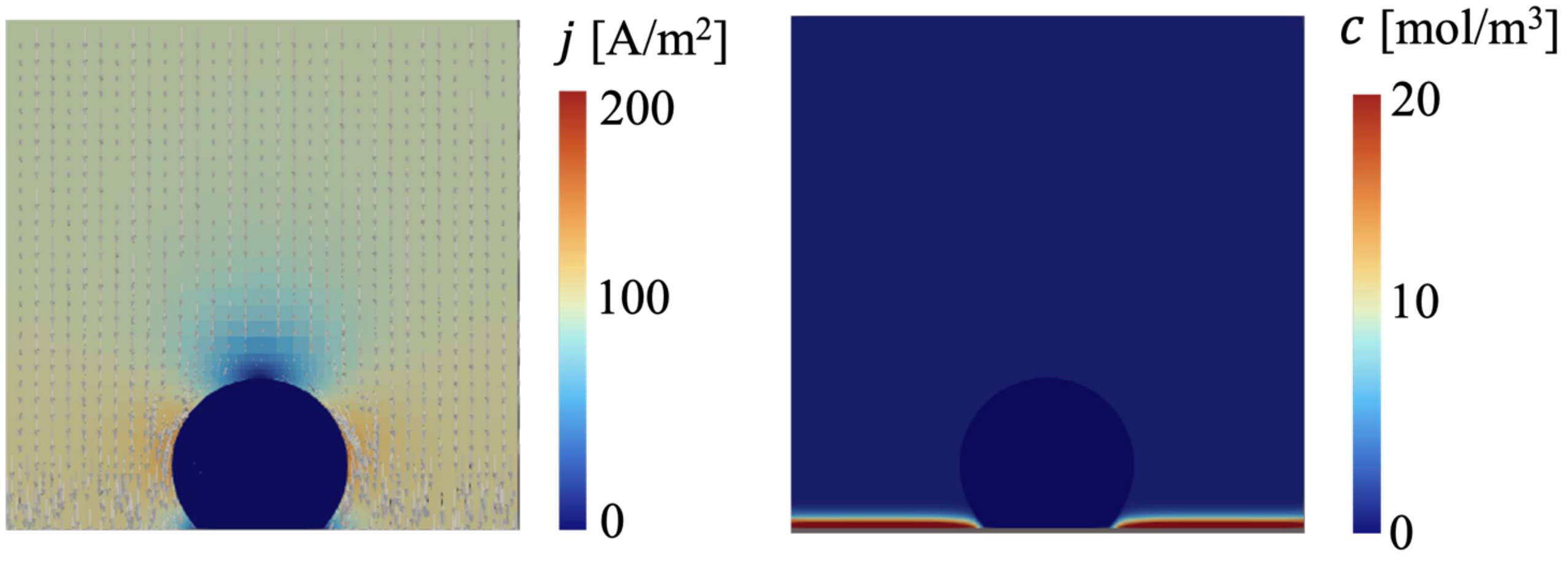

(a) current density [A/m$^2$] (b) gas concentration [mol/m$^3$]

**Fig. 3**. Distribution of (a) the magnitude and direction of the current density and (b) the dissolved gas concentration in the L-e case at $t = 20$ s.

Fig. 4 shows the distribution of the current density and the gas concentration in the S-e case at 20 s. Here, the current density is concentrated at the outer edge of the small disk electrode, a feature well known from the primary current distribution of small disk electrodes (Newman 1966). The dissolved gas concentration closely mirrors the current density pattern. The dissolved gas produced in the wetted ring around the bubble foot is partly absorbed by the bubble, but also partly accumulates in the electrolyte near the electrode.

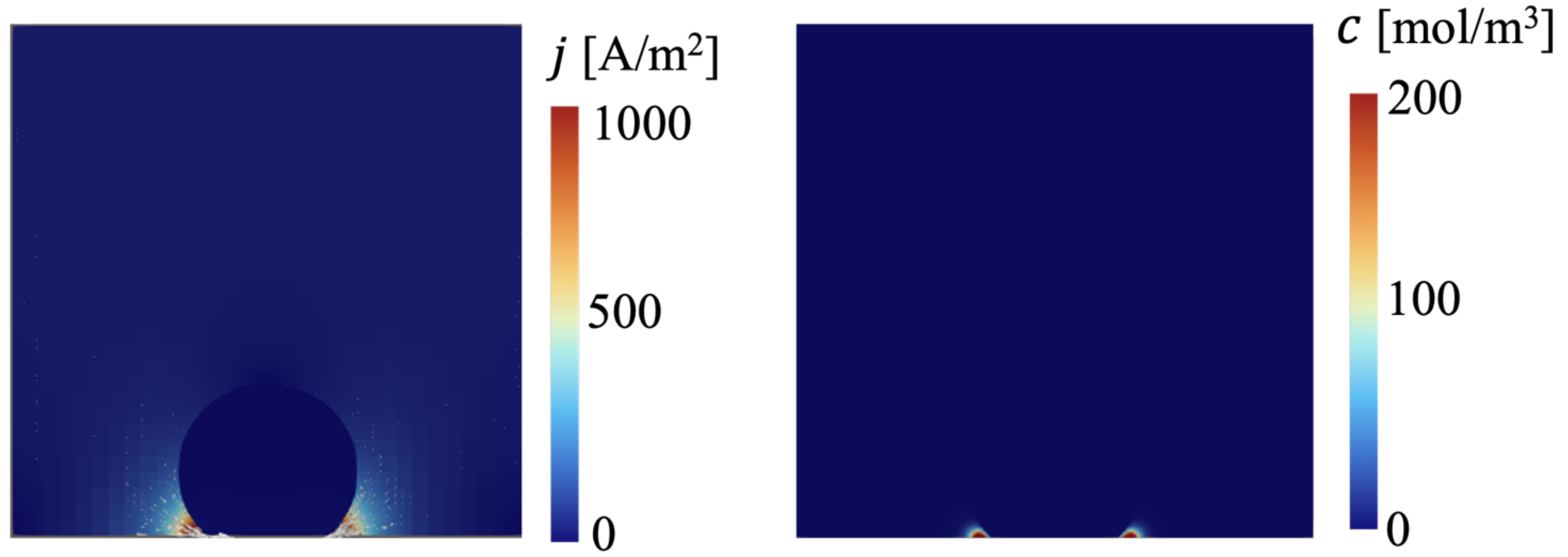

(a) current density [A/m$^2$] (b) gas concentration [mol/m$^3$]

**Fig. 4**. Distribution of (a) the magnitude and direction of the current density and (b) the dissolved gas concentration in the S-e case at $t = 20$ s.

The different electrode sizes lead to different levels of the cell current, as can be seen in Fig. 5(a). The cell current in the L-e case is about three times higher compared to the S-e case owing to the much larger electrode area and thus a lower overall Ohmic resistance. The cell current is more sensitive to the bubble dynamics in the case of the smaller electrode. In the S-e case, the current first decreases, as the growing bubble masks an increasing portion of the disk electrode, but then slightly increases again. As shown in the Appendix (Fig. A2), this later increase is caused by the buoyancy effect, which exerts an upward force on the bubble and thereby reduces the contact area between the bubble and the electrode. In the L-e case, in contrast, the bubble covers only a small fraction of the large electrode, and the cell current is only weakly affected during the entire observation time.

Although the total cell current is higher in the L-e case, the current density averaged over the wetted electrode area in the L-e case (~100 A/m$^2$) is much smaller than that in the S-e case (~1000 A/m$^2$) due to the highly localized current lines at the edge of the small electrode. As a result, gas production is concentrated in the immediate vicinity of the bubble and effectively collected by the bubble in the S-e case, leading to a faster growth, as can be seen in Fig. 5(b). This suggests that electrodes patterned with small catalytic spots, compared to large planar electrodes, could favor fast bubble growth and, subsequently, buoyancy-driven detachment.

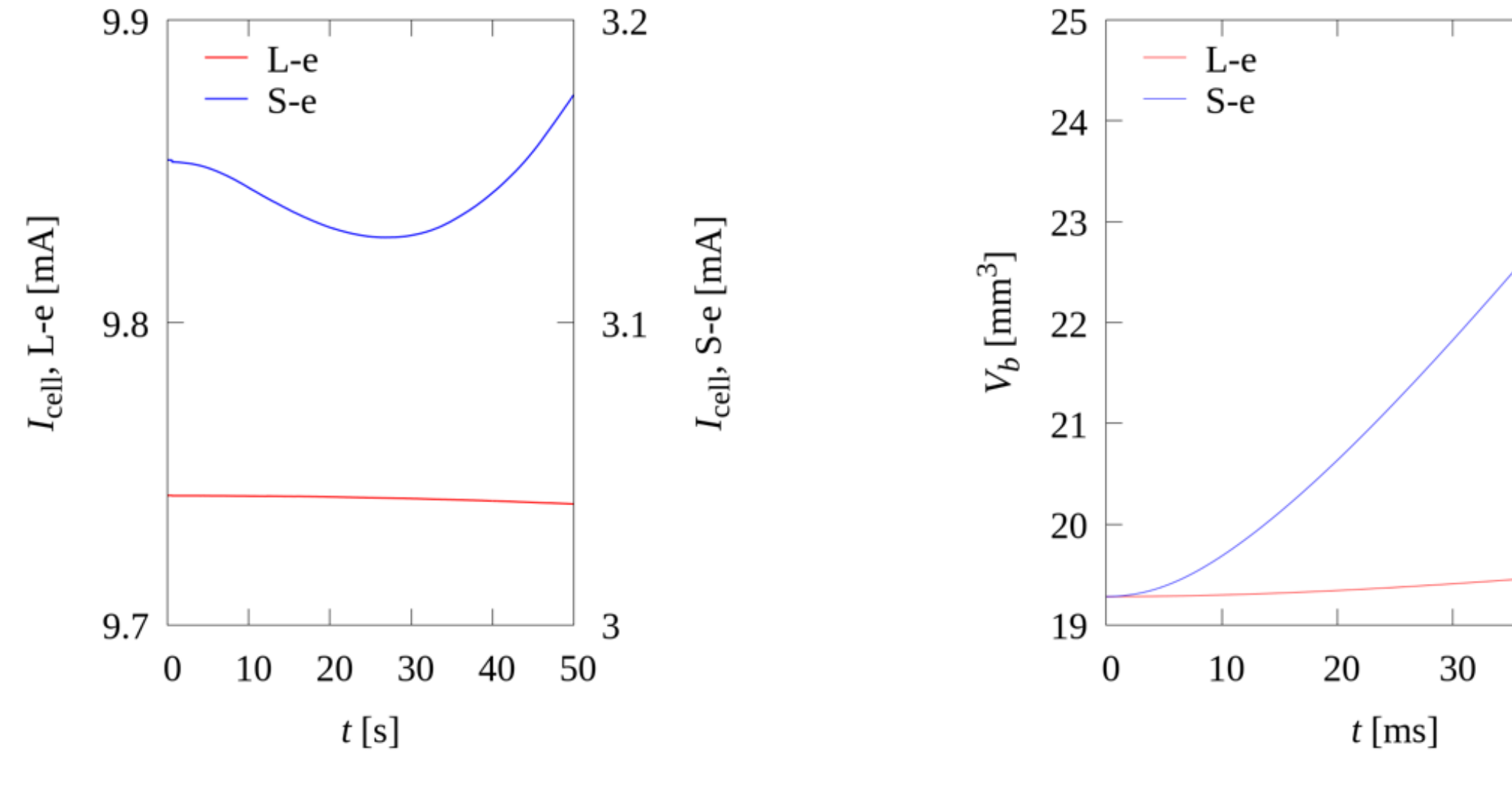

(a) cell current *vs.* time (b) bubble volume *vs.* time

**Fig. 5**. Temporal evolution of (a) the cell current and (b) the bubble volume in the L-e and the S-e case.

### 3.2 Multiple bubbles on a patterned electrode

#### 3.2.1 Influence of bubble number and spacing

Now, we extend the S-e case to multiple bubbles on a patterned electrode. As can be seen in Fig. 6, when the number of the electrocatalytic islands and the distance between the neighbors increase, the distribution of the electric potential on the bottom wall becomes more uniform. This indicates a weaker deflection of the current lines, and thus a lower Ohmic resistance. Initially, the Ohmic resistance in the 1Bub, 3Bub, 9Bub and the 9Bub-Far case with a larger spacing is calculated to be 22.3, 19.2, 18.3 and 15.3 Ω, respectively.

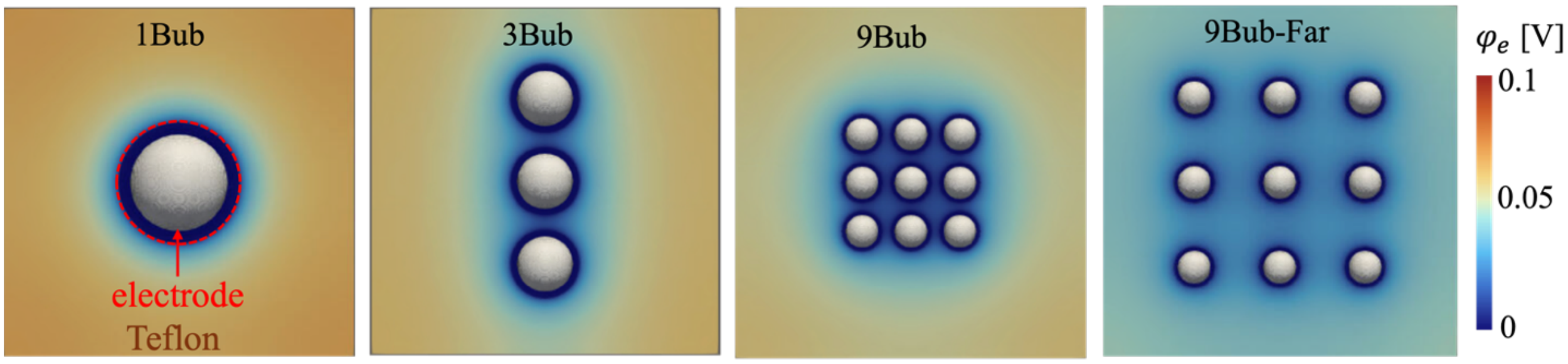

**Fig. 6**. Initial arrangement of the electrode and the bubbles, and the calculated initial electric potential on the bottom wall. The total electrode area and the initial total bubble-electrode contact area are kept the same in all cases. The spacing is 4.2 $r_e$ in the 9Bub-Far case, and 2.4 $r_e$ in other cases.

Corresponding to the initial Ohmic resistance, the initial cell current is larger in the cases of higher bubble number and larger inter-spacing, as can be seen in Fig. 7(a). However, due to the high initial current, the bubbles grow fast and the contact area extends rapidly, causing the cell current to drop quickly. In the 9Bub-Far case, the cell current drops to nearly zero after about 13 s, when the electro-active area is essentially covered by gas. In the case of a single bubble, the cell current starts from a relatively low value and decreases only moderately. In the other multiple-bubble cases (3Bub, 6Bub and 9Bub), the cell current decreases continuously with the increasing gas-solid contact area during the observation time.

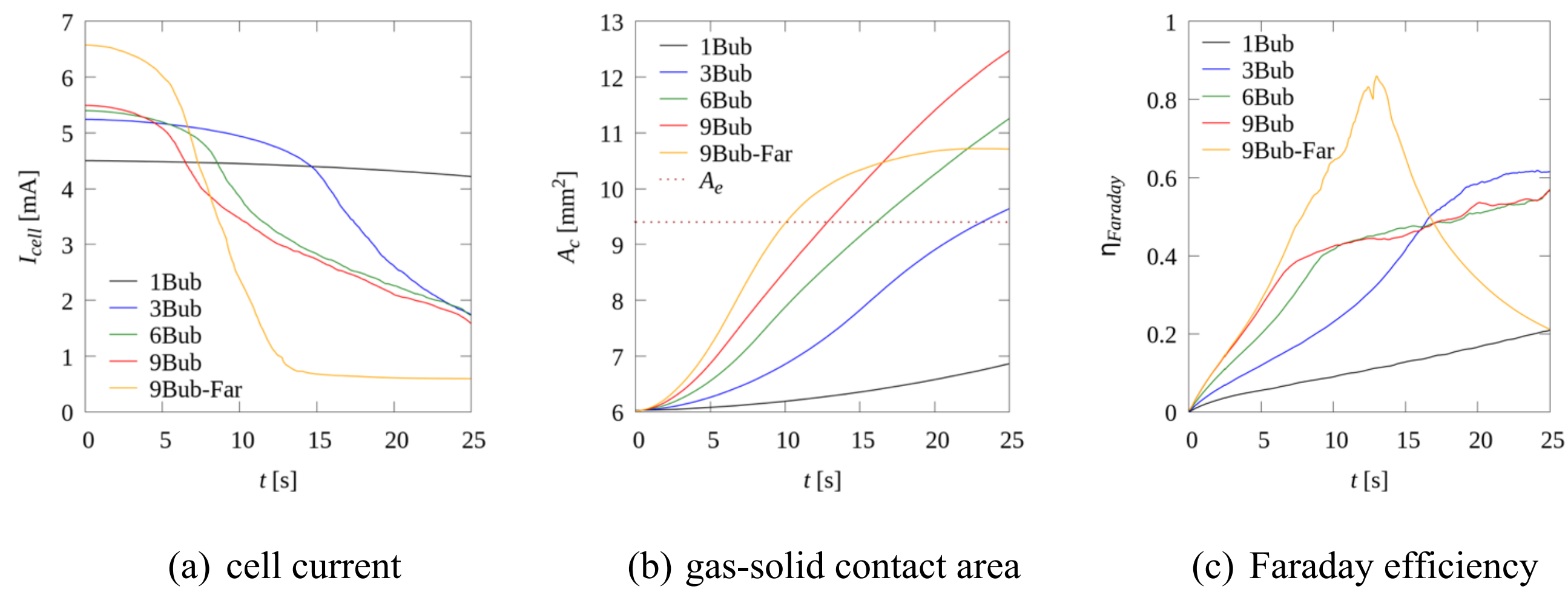

(a) cell current (b) gas-solid contact area (c) Faraday efficiency

**Fig. 7**. Temporal evolution of (a) the cell current, (b) the gas-solid contact area and (c) the Faraday efficiency, Eq. (13), in the multiple bubble cases.

We further evaluate the Faraday efficiency, defined in Eq. (13), which measures the fraction of the produced gas that actually enters the bubbles, instead of remaining dissolved in the electrolyte. As can be seen in Fig. 7(c), for the case 9Bub-Far, $\eta_{Faraday}$ first increases rapidly, but at about 13 s when the electrode area is completely covered by the bubbles it continuously decreases again. In other cases, $\eta_{Faraday}$ increases with time, as the wetted area is shrinking and more dissolved gas is directly feeding the bubble rather than remaining in the liquid. In general, a larger $\eta_{Faraday}$ is found when there are more bubbles and when the inter-spacing is larger, except in the later stage when the electrode has become almost entirely covered by the bubbles.

$$\eta_{Faraday} = \frac{dV}{dt}\frac{zF\rho_g}{I_{cell}M_g} \quad \text{Eq. (13)}$$

We further notice in Fig. 7(b) that the gas-solid contact area can exceed the entire electrode area in the multiple-bubble cases, while the cell current has not yet reached zero. To understand this, we focus on the 3Bub case and investigate the distribution of the gas fraction and the current density at the bottom wall of the domain. As can be seen in Fig. 8, initially, the bubbles (magenta lines) are located at the center of their corresponding electrodes. The current density shows a slight asymmetry, with a higher value on the outer side of the outer electrodes due to the edge effect. As the reaction proceeds and the bubbles grow, the two outer bubbles are found to move outwards during growth, covering the outer sides of the outer electrodes completely, but keeping the inner sides of the outer electrodes wetted, where the current density then localizes. Accordingly, the dissolved gas also accumulates on the inner sides of the outer electrodes. This outward shift of the outer bubbles is the reason why the cell current remains non-zero although the total gas-solid contact area already far exceeded the entire electrode area. In the following, we will analyze the reason causing this bubble drift in detail.

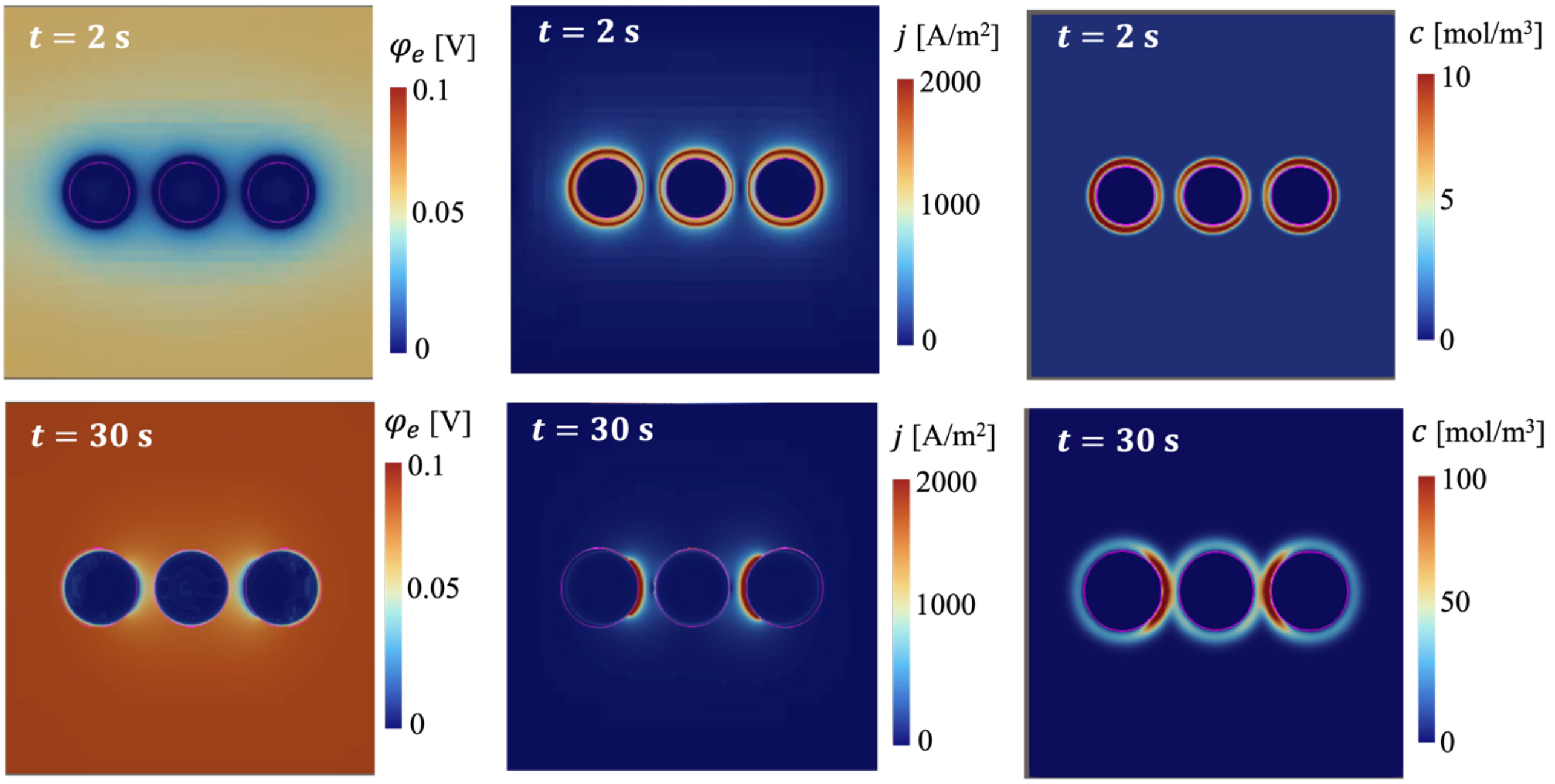

**Fig. 8**. Distribution of the electric potential (left), the current density (middle) and the dissolved gas concentration (right) at 2 s (top) and 30 s (bottom) in the case of three electrode-bubble pairs. The magenta lines represent the bubble contours.

### 3.2.2 Bubble drift

As can be seen in Fig. 8, the current density is initially more concentrated on the outer sides of the two outer electrodes. This may result in a higher local gas production rate, which may cause the bubbles to grow towards the outer direction. To check whether this asymmetry is the dominant cause of the drift, we apply a constant current density on the wetted part of the electrode instead of applying a constant potential boundary condition. The details of the results are given in the Appendix. As can be seen in Figs. A3-A4, the outer bubbles drift outwards also when a uniform current density is applied, indicating that the current-density asymmetry alone cannot explain the drift.

Another possible reason for the bubble drift is related to the growth-driven advection: each growing bubble displaces the surrounding liquid radially, and a neighboring bubble immersed in this flow is pushed away. The incompressibility condition requires the radial velocity to decay with the squared distance from the bubble center, as in the kinematics underlying the Rayleigh-Plesset equation (Plesset and Prosperetti 1977). Balancing the volume flux across the surface of the growing central bubble with that of a concentric spherical surface reaching the inner side of the outer bubble, the resulting drift velocity can be estimated by:

$$u_d = \frac{\frac{r_b^2 dr_b}{dt}}{d^2}. \quad \text{Eq. (14)}$$

Here, $r_b$ represents the bubble radius, $d$ represents the center distance between the neighbor electrodes minus $r_b$, so that $u_d$ corresponds to the velocity at the inner side of the outer bubbles, which dominates the drift. Applying Faraday's law and assuming an ideal gas, we can further simplify Eq. (14). The change of bubble volume with time can be expressed as:

$$\frac{dV_b}{dt} = \frac{\eta_{Faraday} I_{cell} M_g}{zF\rho_g}. \quad \text{Eq. (15)}$$

Here, $\eta_{Faraday}$ is the Faraday efficiency. $M_g, \rho_g, z, F$ are the molar mass and density of the gas, the charge number and the Faraday constant. Assuming that the bubbles maintain a hemi-spherical shape, Eq. (15) is reformulated:

$$\frac{d\left(\frac{2n}{3}\pi r_b^3\right)}{dt} = \frac{\eta_{Faraday} I_{cell} M_g}{zF\rho_g}, \quad \text{Eq. (16)}$$

with $n$ representing the number of electrode-bubble pairs.

Combining Eqs. (14) and (16), we then obtain:

$$u_d = \frac{\eta_{Faraday} I_{cell} M_g}{2n\pi zF\rho_g d^2}. \quad \text{Eq. (17)}$$

As can be seen, if $\eta_{Faraday}$ and $I_{cell}$ are constant values, the shift velocity should also remain constant. To clarify the mechanism causing the bubble drift, Fig. 9 shows how the bottom centers of the three bubbles move with time, overlaid by the drift predicted from Eq. (14) using the simulation data of $r_b$, as well as the analytical solution Eq. (17) with $\eta_{Faraday}$ and $I_{cell}$ given from the simulation data averaged between 0 and 25 s, see Fig. 7. The good agreement between numerical and theoretical results indicates that the growth-driven electrolyte flow between the neighboring bubbles should be the dominant mechanism of the drift.

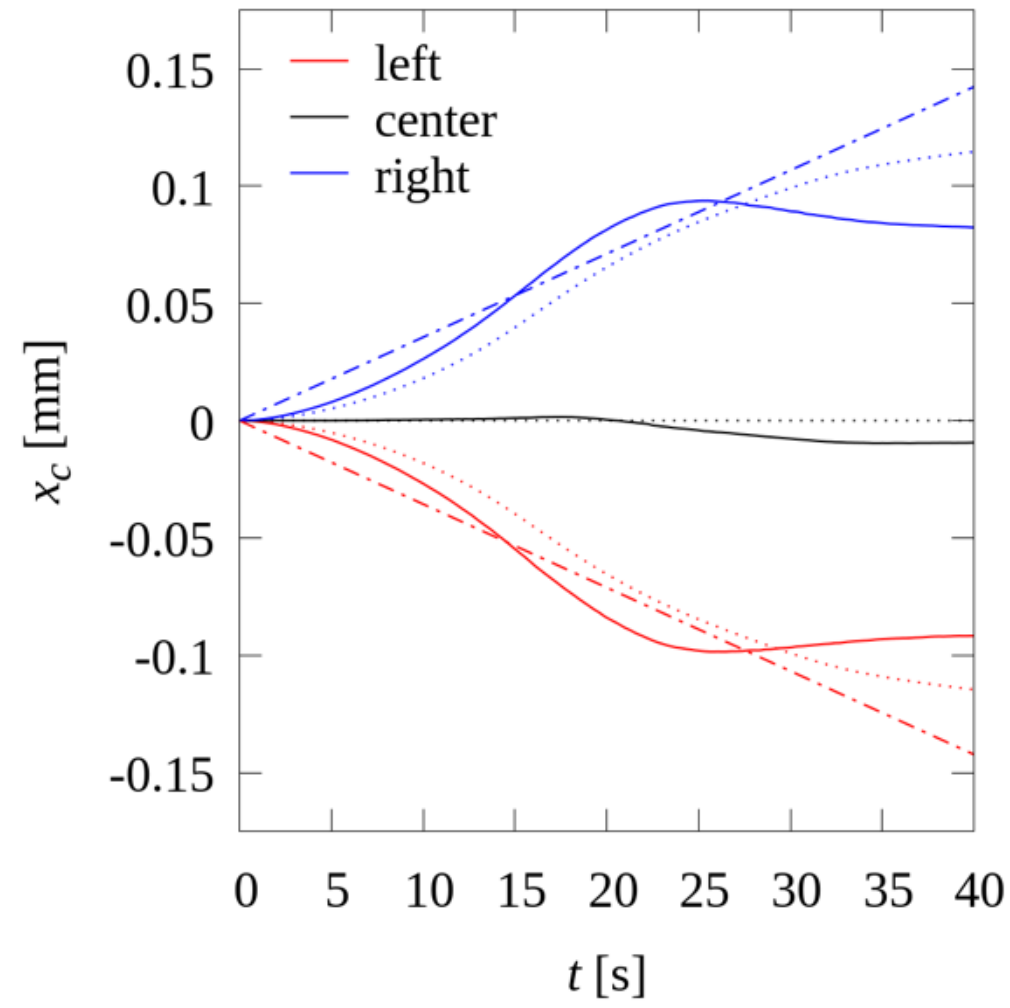

**Fig. 9**. Simulated (solid lines) and theoretically predicted bubble drift using Eq. (14) (dotted lines) and Eq. (17) (dash-dotted lines) for the case with three bubbles. To suppress numerical noise, $r_b(t)$ is smoothed with a moving average before evaluating $dr_b(t)/dt$ in Eq. (14).

Figure 10 visualizes the flow field around the three bubbles and the horizontal velocity profile along a line crossing the bubbles. As can be seen, the two outer bubbles are pushed outwards, and the flow velocity is quantitatively in line with the outward drift in Fig. 9: the simulated velocity magnitude of ~0.003 mm/s agrees well with the theoretical prediction.

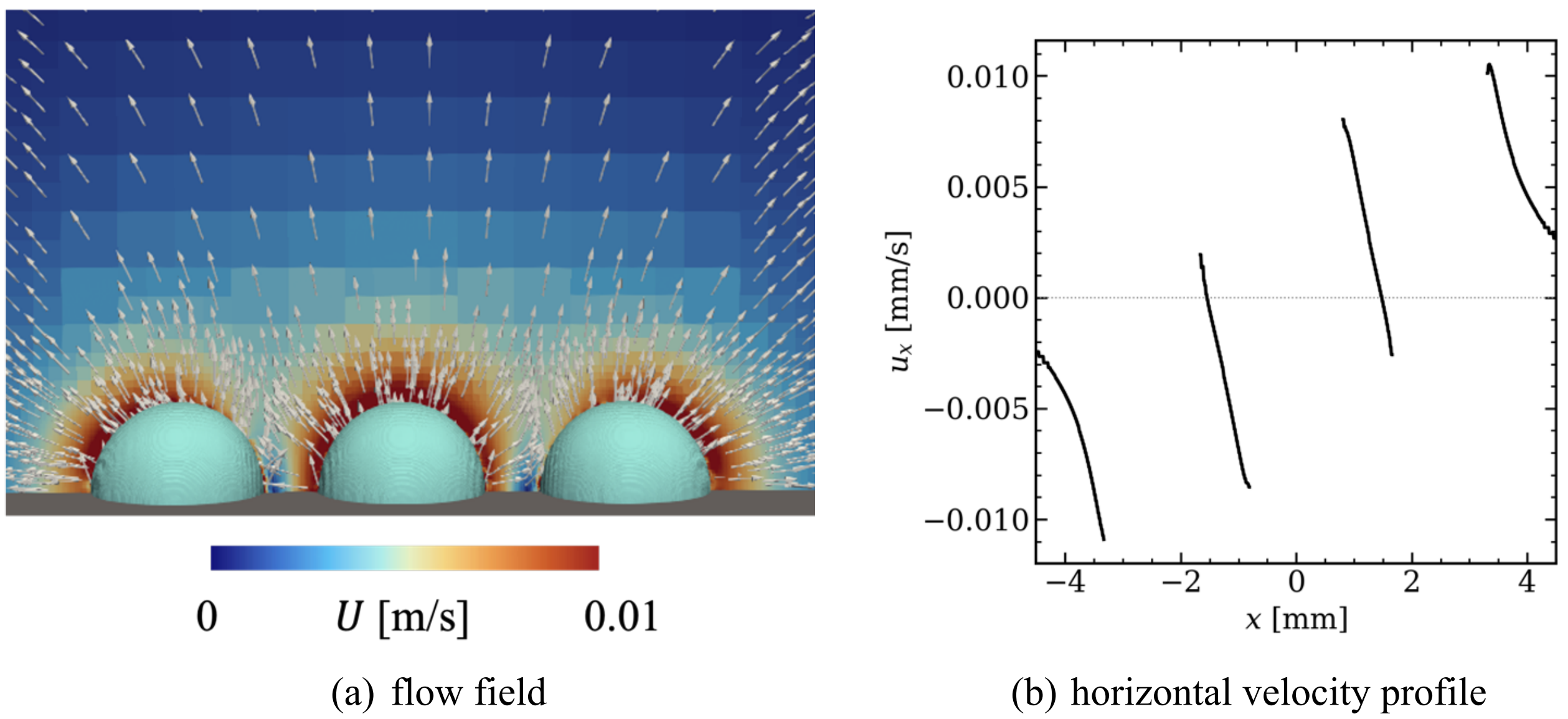

**Fig. 10**. (a) Flow field at 20 s. Color surface: magnitude of the flow velocity. Arrows: Flow vectors. (b) Horizontal velocity $U_x$ along a horizontal line crossing the three bubbles and 0.5 mm (~50% $r_b$) above the bubble bottom at 20 s.

Further, we analyze how the electrode spacing influences the bubble drift in the three-bubble cases. Fig. 11(a) shows the drift when the distance between neighboring electrodes increases from 2.1 $r_e$ to 3 $r_e$. We remark that, $r_e - r_b = 0.2$ mm should be an upper limit of the drift, which is obtained by assuming that the bubble has not grown and has moved from the center of a disk electrode to the position that will cross the electrode. After that, the outer side of the electrode will be completely covered by the bubble, and the asymmetry of gas production on the electrode is expected to become more dominant, which would slow down or even reverse the outer drift of the bubble. The later reversal of the outer drift can be observed in simulations especially when the spacing between neighbor bubbles is large. In general, a larger drift is observed for smaller spacing. The theoretical prediction based on the advection-driven drift (Eq. 14) shows the same order of magnitude as the simulations for the different spacings, but the later reversal of the drift, which is caused by the asymmetry of the current density, is not captured by the theory.

Assuming that the drift is dominated by the growth-driven advection until reaching its peak, the peak value of the drift should be an indicator of the strength of the advection. We plot in Fig. 11(b) the peak drift versus $1/d^2$, and found a linear relation. This further supports the theoretical analysis of the advection-driven drift (Eq. 14 or 17).

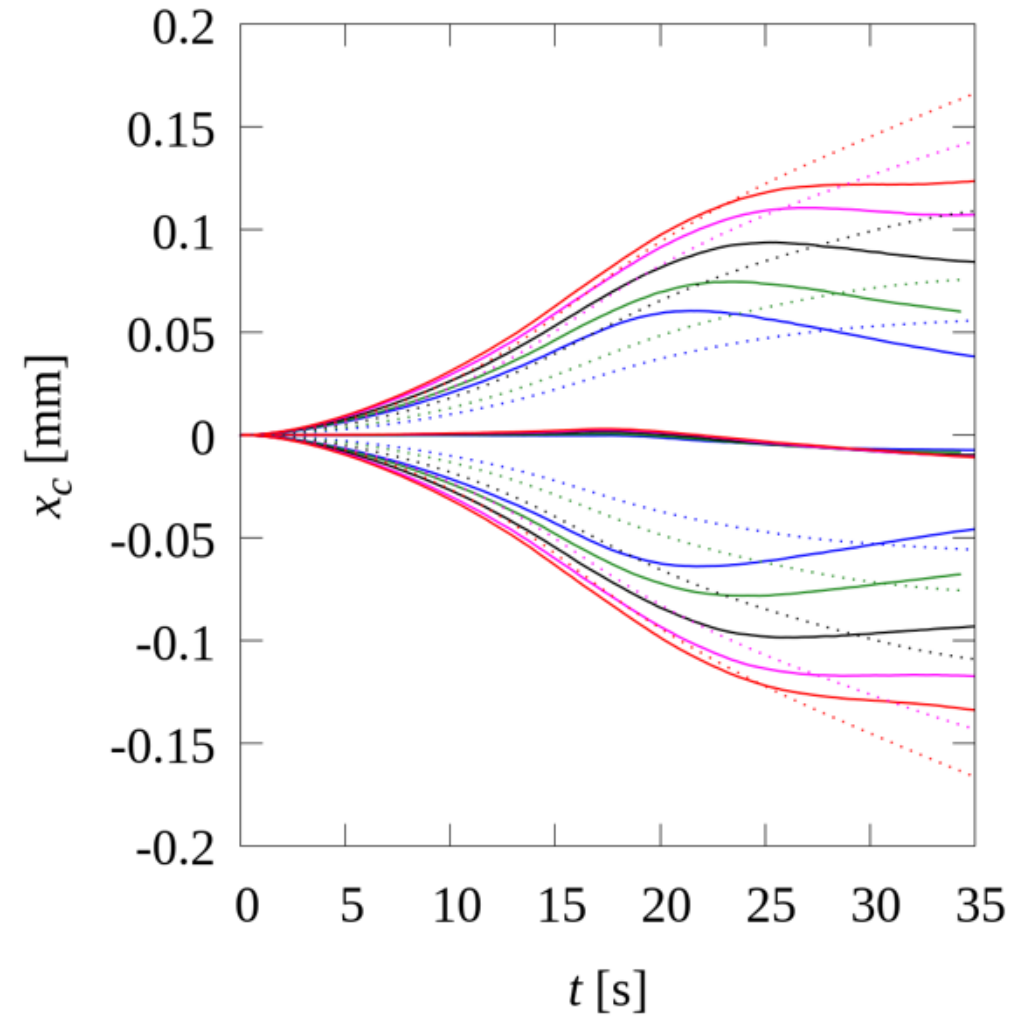

(a) bubble drift *vs.* time

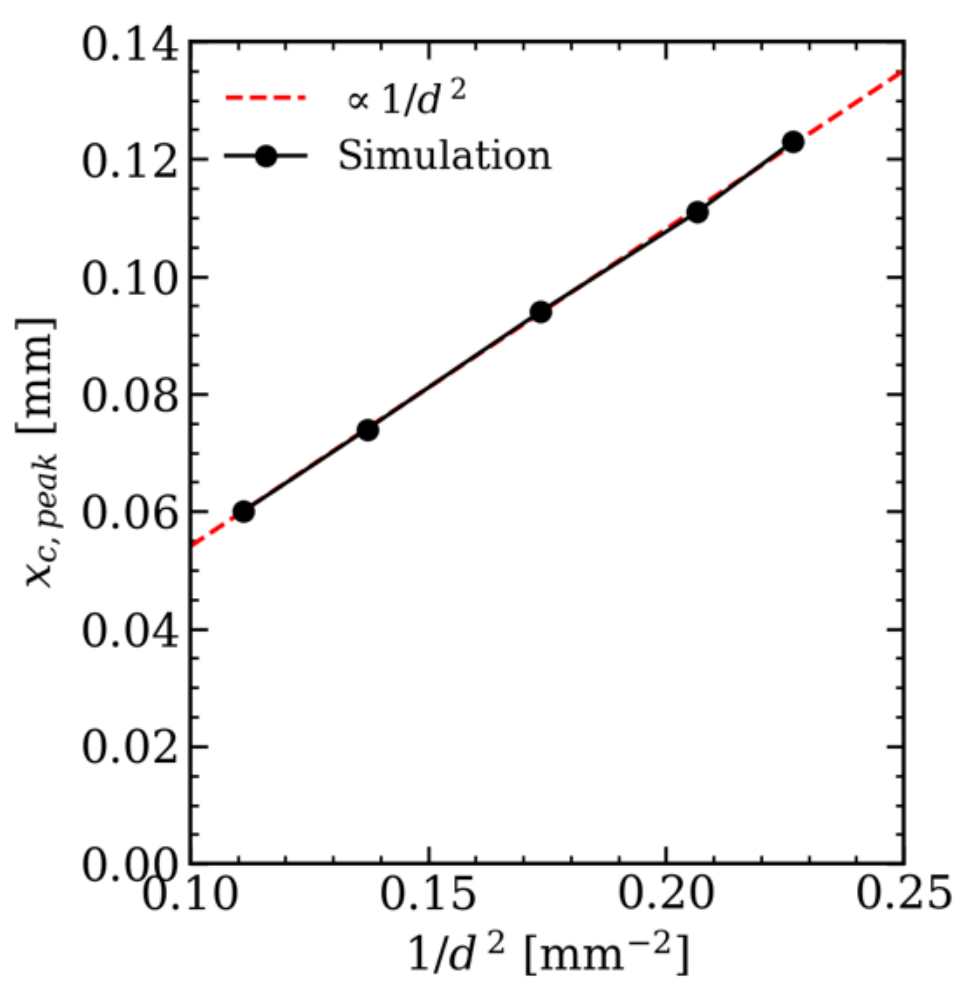

(b) drift peaks *vs.* inter-bubble spacing

**Fig. 11**. (a) Influence of spacing on the temporal behavior of bubble drift in the three-bubble cases. The distance between the neighbor electrodes corresponds to 2.1 (red), 2.2 (magenta), 2.4 (black), 2.7 (green) and 3 (blue) times electrode radius. The dotted lines are predicted values from Eq. (14).

To suppress numerical noise, $r_b(t)$ is smoothed with a moving average before evaluating $dr_b(t)/dt$ in Eq. (14). (b) The maximal drift as a function of $1/d^2$.

Overall, the influence of the spacing is two-fold. On one hand, a larger spacing reduces the Ohmic resistance and yields a higher initial current, as can be seen in Fig. 12(a). As the bubble continues to grow, the outer bubbles start to drift outwards, which is stronger at smaller spacing. This postpones the complete coverage of the electrode and keeps the cell current at a moderate level for a longer time. These two competing effects are summarized in Fig. 12(b), which shows the total transferred charge $Q = \int I_{cell} dt$ as a function of the spacing, evaluated during two time windows: In the early stage (0-15 s), a larger spacing produces a more uniform electric field and thus a lower Ohmic resistance, allowing more charge to be transferred. During the entire observation time (0-35 s), this trend reverses, as at later times the outward drift of non-central bubbles becomes more dominant, so that a smaller spacing transfers more charge.

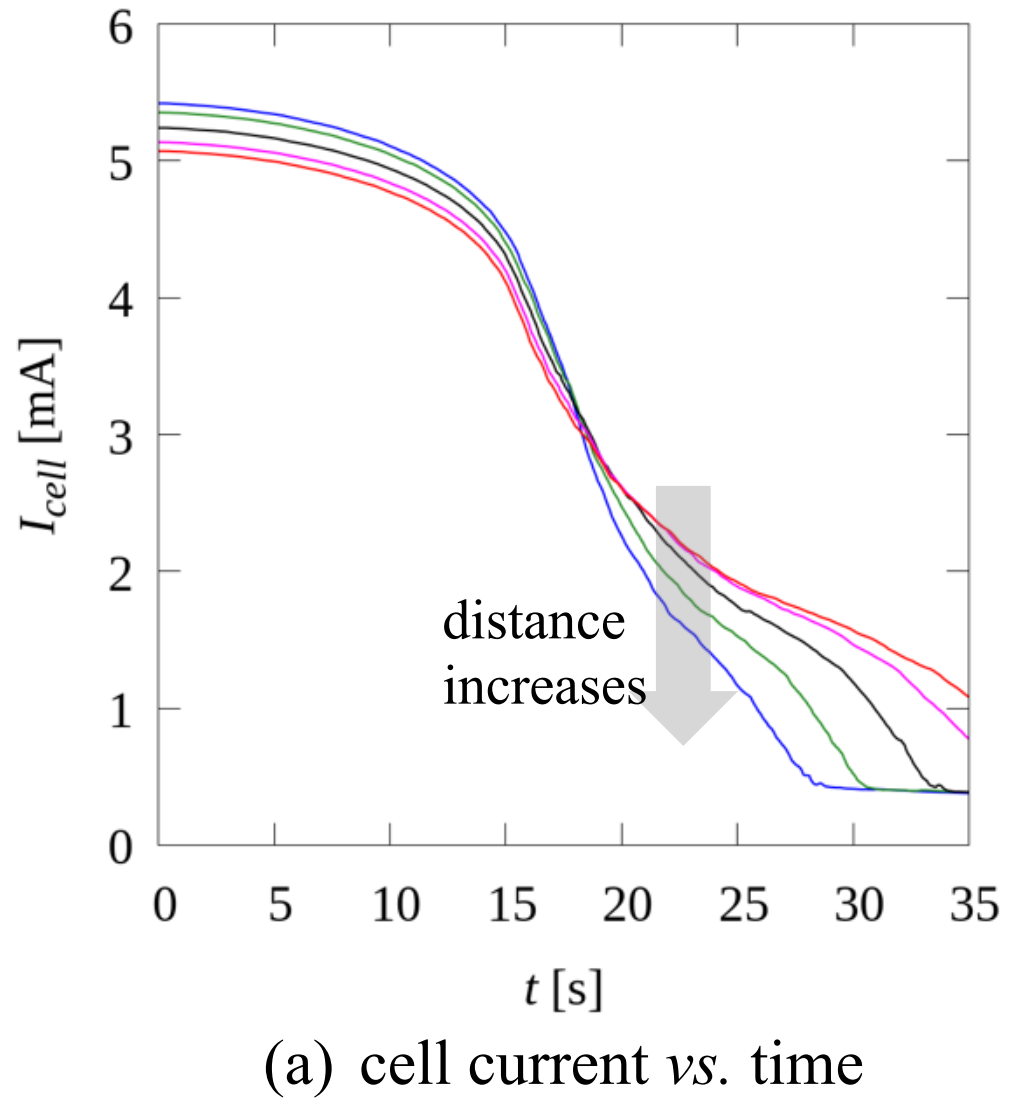

(a) cell current *vs.* time

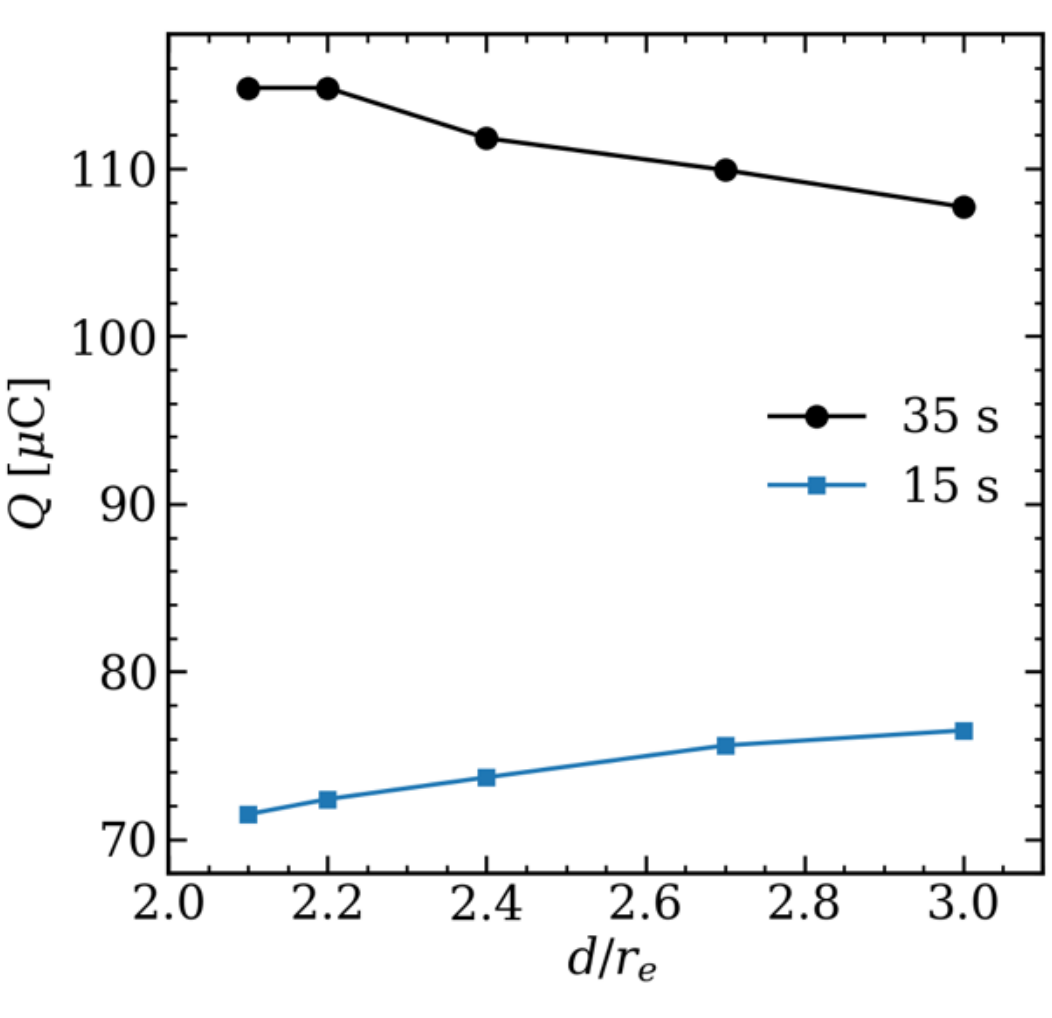

(b) transferred charge *vs.* spacing

**Fig. 12**. (a) Influence of spacing on the cell current in the three-bubble cases. The distance between the neighbor electrodes corresponds to 2.1 (red), 2.2 (magenta), 2.4 (black), 2.7 (green) and 3 (blue) times electrode radius. (b) The transferred charge in 15 s and in 35 s depending on the spacing.

## 4 Conclusions and Outlook

In this work, the primary electric field is implemented in the electrochemical framework in Basilisk. Going beyond the commonly used assumption of a uniform current density at the wetted part of the electrode, the present simulations capture how the current density distribution and the bubble dynamics influence each other, from a single bubble on electrodes of different size to bubble arrays on a patterned electrode composed of small catalytic islands. Main conclusions include:

(1) At the same cell voltage, a bubble grows faster on a small disk electrode compared to a large planar electrode, as the current density concentrates at the edge of the small electrode.

(2) A patterned electrode with more and smaller catalytic islands results in a smaller overall Ohmic resistance and a higher initial cell current. However, the bubbles then grow fast and quickly cover the electro-active areas, resulting in a rapid decrease of the current.

(3) In bubble arrays, the outer bubbles could drift outwards during the growth, keeping the inner sides of the outer electrodes wetted and thereby delaying the complete coverage of the electrode and sustaining the cell current.

(4) The bubble drift is dominated by the growth-driven advection and scales with $1/d^2$.

(5) The spacing of the catalytic islands emerges as a design parameter that balances a low initial Ohmic resistance against a delayed gas coverage of the electrode.

We have to admit, that the results may be quantitatively influenced by the simplified wetting modeling of a constant contact angle that is identical for the microelectrode and the neighboring passive wall, thereby excluding pinning effects. In future work, more advanced wetting models will be used, and the present framework will be extended to secondary and tertiary current distributions accounting for reaction kinetics and concentration overpotential. Finally, also a direct comparison with experiments on patterned electrodes will be targeted.

**Appendix**

Fig. A1 shows how the cell current evolves with time in the M-e case with three bubbles when the electric conductivity of the gas phase is set to be 0.1 S/m or 0.01 S/m. In general, the results are very similar. Only at a later stage, when most parts of the electrode have been covered by the bubbles, a lower $\kappa_g$ leads to lower cell current approaching zero. To avoid the high computational efforts at small $\kappa_g$ while still ensuring accurate results, we choose $\kappa_g = 0.1$ S/m in our simulations.

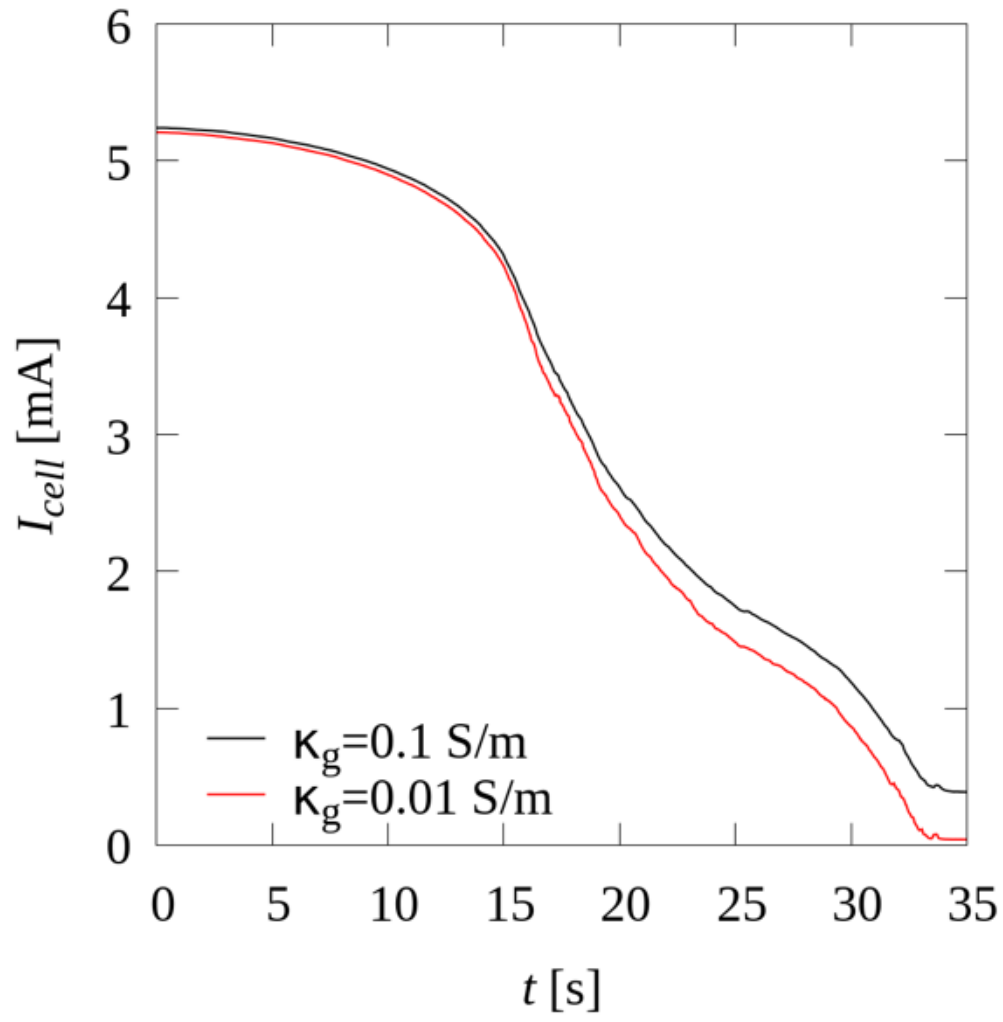

**Fig. A1**. Temporal evolution of the cell current in the three-bubble case when the electric conductivity in the gas phase is set to be 0.1 S/m and 0.01 S/m.

Fig. A2 shows how the gas-solid contact area and the cell current change with time when the gravitational acceleration is activated and disabled in the simulations of the S-e case. If there is no gravity or the buoyancy effect, the bubble coverage on the electrode increases with the bubble growth, causing the cell current to reduce greatly. With the gravity, the upward buoyancy force exerts on the bubble and avoids the increase of the contact area, keeping the cell current at a high level.

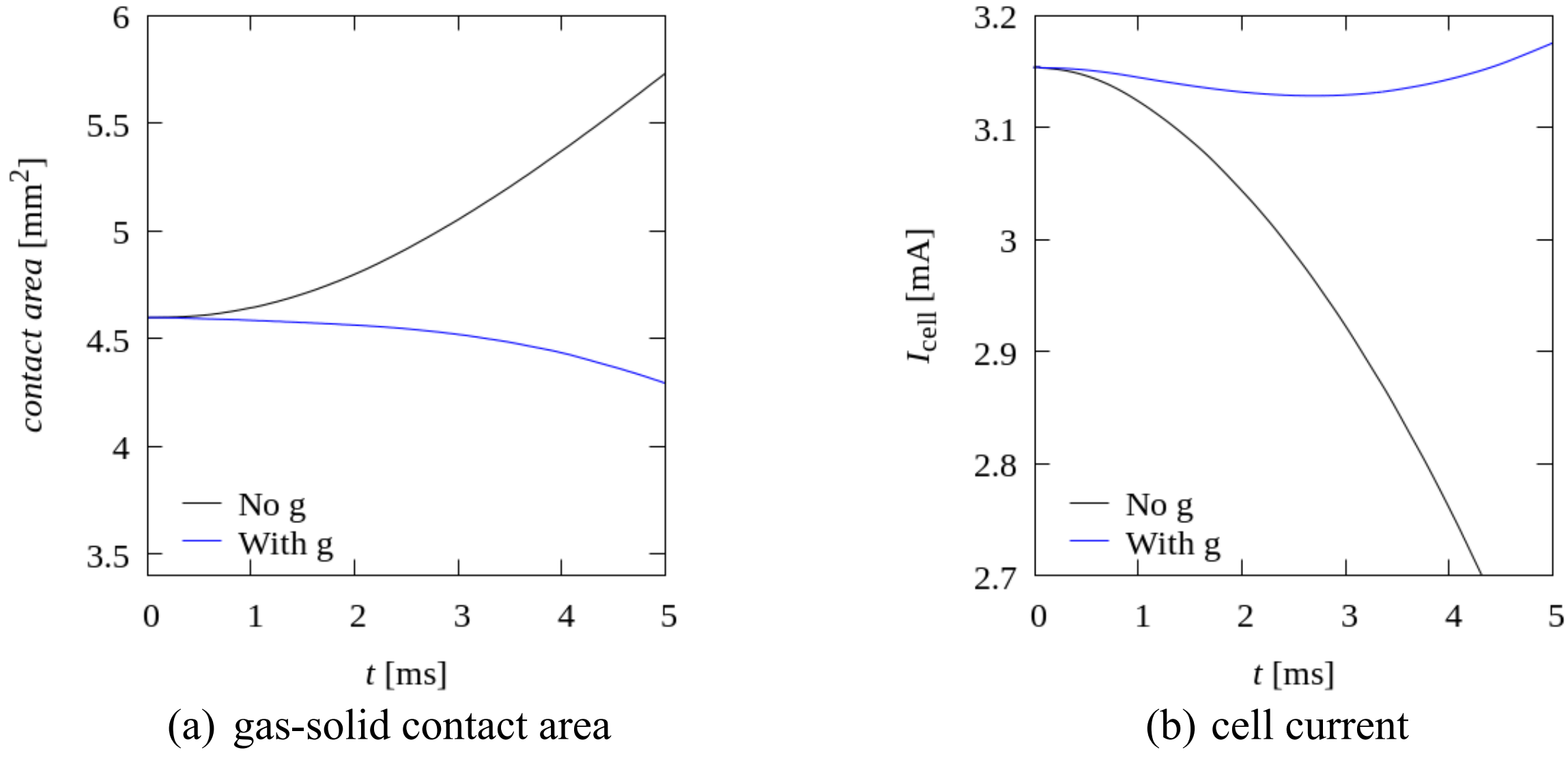

(a) gas-solid contact area (b) cell current

**Fig. A2**. (a) gas-solid contact area and (b) cell current with and without gravity in the S-e case.

As can be seen in Fig. 8, the current density is initially more concentrated on the outer sides of the outer two electrodes. This may result in higher local gas production rate, which may also cause the bubbles to grow towards the outer direction. To observe influence of the initial non-uniform distribution of the current density, we investigate how the bubble position evolves when the electric field is not resolved and a constant current density of 1860 A/m$^2$ (corresponds to the average value for a cell voltage of 0.1 V) is applied to the wetted part of the electrode. As can be seen in Fig. A3, the outer bubbles also shift outwards with time, causing accumulation of dissolved gas on the inner, uncovered parts of the electrodes. Fig. A4 compares temporal evolution of the bottom center of the bubbles when the electric field is resolved ($j_{var.}$) or not resolved ($j_{cons.}$). At the similar level of the applied current density, the magnitude of the bubble shift is also similar. As $j_{cons.}$ is larger than $j_{var.}$ at the beginning and smaller than $j_{var.}$ later, the bubble shift is also first faster, then slower compared to the case of $j_{var.}$

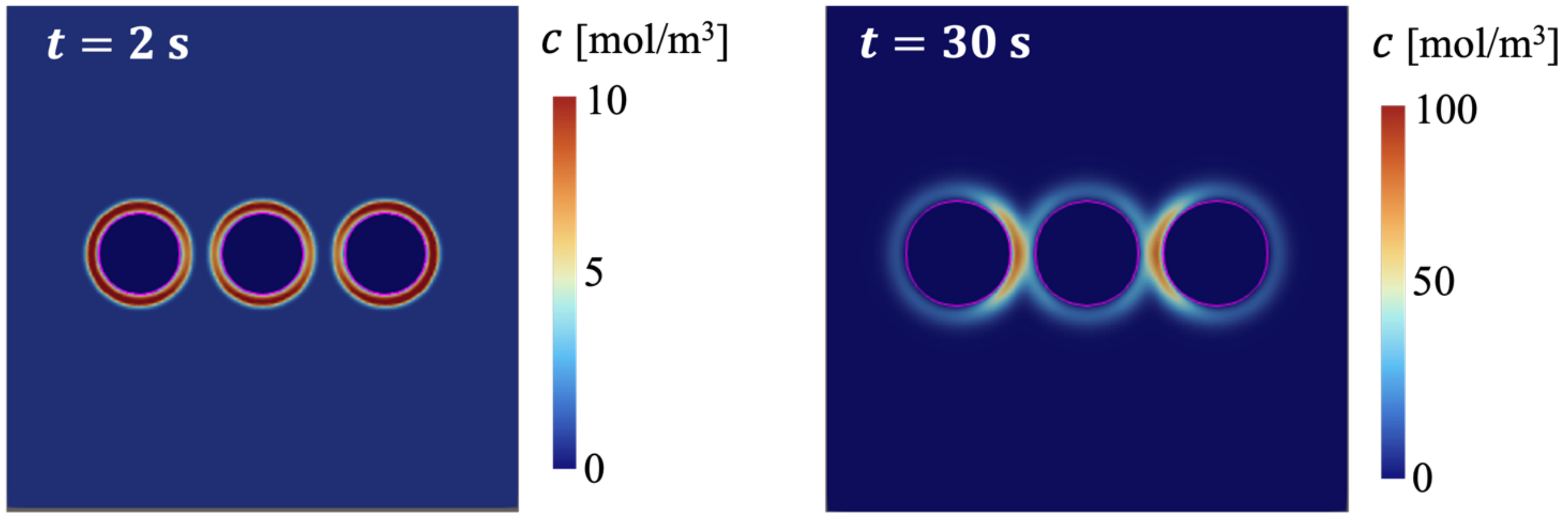

**Fig. A3**. Distribution of the dissolved gas concentration at 2 s and 30 s when the electric field is not resolved and a constant current density is applied on the wetted part of the electrode. The magenta lines represent the bubble contours.

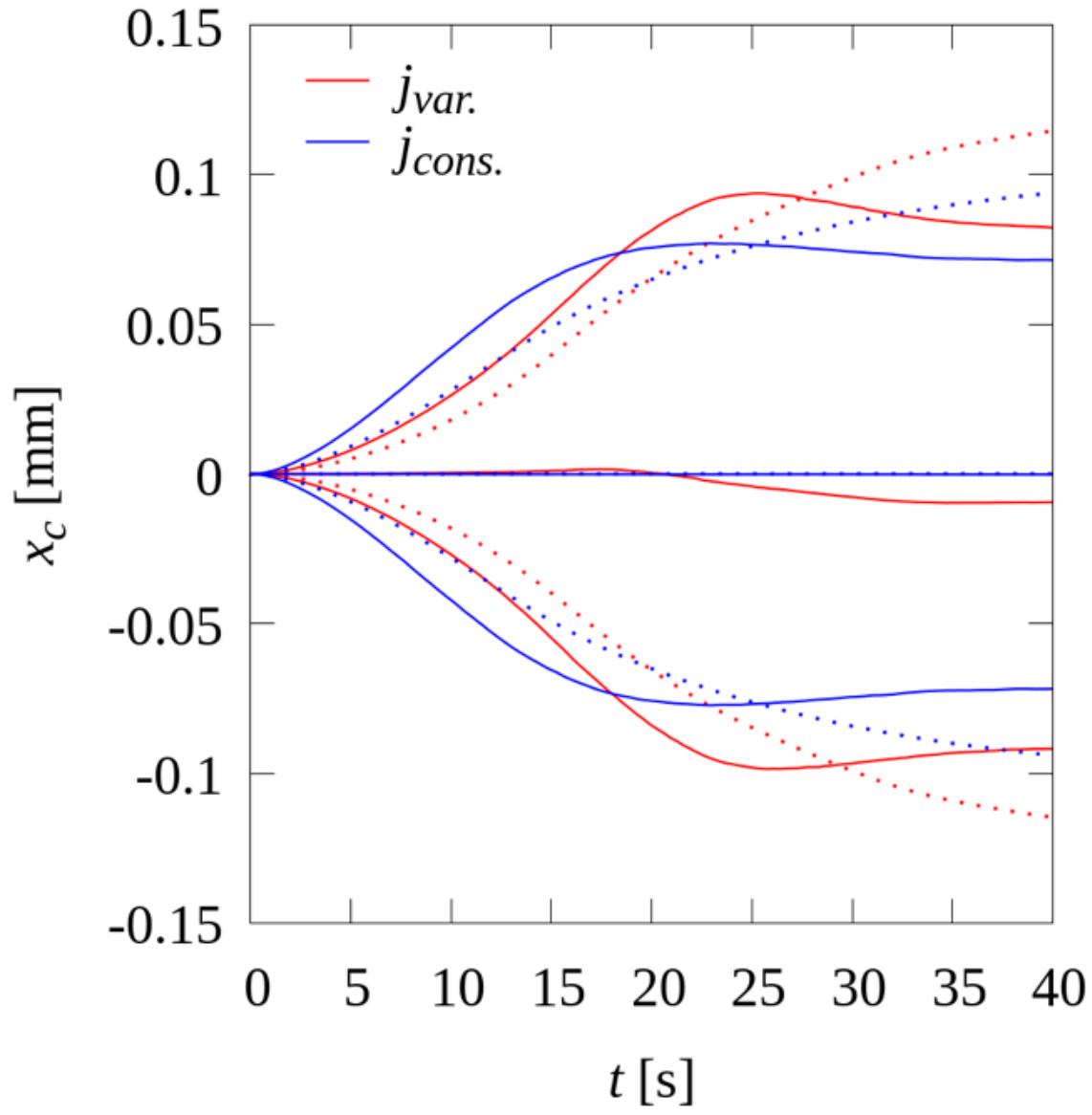

**Fig. A4**. Simulated (solid lines) and theoretically predicted bubble drift using Eq. (14) (dotted lines) for three bubbles when the electric field is resolved ($j_{var.}$) and when a constant current density ($j_{cons.}$) is applied. To suppress numerical noise, $r_b(t)$ is smoothed with a moving average before evaluating $dr_b(t)/dt$ in Eq. (14).

**Acknowledgements**

We would like to thank Stéphane Zaleski, Yifan Han, Wei Qin, Tianyang Han and Jieyun Pan for fruitful discussions. We gratefully acknowledge financial support by the National Natural Science Foundation of China, 22309007 and by the German Federal Ministry of Research, Technology and Space within the project ECCM-ALKALIMIT, grant no. 03SF0731B. The computations were performed on the cluster "Rosi" at Helmholtz-Zentrum Dresden-Rossendorf.

**Declaration of competing interest**

The authors have no competing interests to declare that are relevant to the content of this article.